# Comparison of cosmic ray flux at $\sqrt{s} \geq 14$ TeV with LHC luminosity

Frank E. Taylor
MIT
Department of Physics and Laboratory of Nuclear Science
77 Massachusetts Avenue, Cambridge, MA 02139 USA
April 14, 2008
fet@lns.mit.edu

The high energy cosmic ray flux impinging on the sun and earth for 4 Gyr is compared to the operation of the CERN Large Hadron Collider (LHC) at design energy and luminosity. It is shown by two different calculations that both the integrated luminosity and the total hadronic interaction rate from the cosmic ray flux of comparable energy are many orders of magnitude larger than that of the LHC operated for 10 years. This study indicates that it is extremely unlikely that pernicious exotic particles, such as mini-black holes, would be produced by the LHC that would destroy the earth.

## INTRODUCTION

Of recent excitement are speculative models that predict copious black hole production at the LHC. In these theories mini-black holes (BHs) are produced when partons interact at separations closer than the Schwarzschild radius. The BHs will decay into distinctive hadron jets and high $P_T$ leptons allowing BH events to be isolated from background in a very short time. One such model is that of Arkani-Hamed, Dimopoulos, and Dvali [1] in which the solution of the so-called hierarchy problem is accomplished by postulating extra (large) spatial dimensions in a theory with just one energy scale, of order one TeV, which not only accommodates electroweak scale breaking but also unifies strong, electroweak and gravitational forces. Gravity, in this model, is weak because its wave functions spread in the extra dimensional 'bulk' where they have little overlap with the Standard Model particles such as fermions and bosons that 'live' in conventional 4-dimensional space.

It has been suggested that the mini-black holes or some other pernicious exotic particles produced at the LHC would have an adverse effect on the habitability of the earth [2]. While such processes cannot be absolutely precluded on first principles, this note shows that if such an effect exists, energetic cosmic rays impinging on either the sun or the earth would have provided the 'excitement' of nearby black holes eons ago. Thus, our very existence precludes this extremely unlikely 'doomsday'.

Nevertheless, it is of interest to compute how dangerously close we have come to this putative mass destruction. In this note, we compute the flux and interaction rates of cosmic rays which have an equivalent center of mass energy or higher impinging on the sun and earth over their respective lifetimes. It is shown by two different calculations based on the lifetime of the sun and earth that the probability is vanishingly small for such a destructive event to occur at the LHC.

The integrated cosmic ray flux and target densities of the sun and earth are the key factors in these estimates. In addition to the overwhelmingly large cosmic ray flux and cumulative interactions versus that which will be provided by the LHC, it is shown that if black holes were to be produced, they would decay by the Hawking Mechanism [3] extraordinarily rapidly and would not be able to express their malignant tendency to 'eat' matter.

## LHC EQUIVALENT ENERGY OF COSMIC RAYS

The cosmic ray flux has a rapidly falling power-law spectrum which extends up to greater than $10^{20}$ eV ($10^{11}$ GeV) – about the kinetic energy of a well-hit tennis ball ($10^{20}$ eV ~ 20 joules) [4]. The LHC collider will operate at $\sqrt{s}$ = 14 TeV. The equivalent cosmic ray energy impinging on a fixed proton (nucleon) target is determined by:

$$s = (14\ \text{TeV})^2 = (2E_b)^2 \sim 2E_cM_p, \tag{1}$$

where $E_c$ is the incident cosmic ray energy, $M_p$ is the proton rest mass and $E_b$ is the LHC beam energy (7 TeV). Thus,

$$E_c \sim 1.0\text{x}10^8\ \text{GeV} = 1.0\text{x}10^{17}\ \text{eV}. \tag{2}$$

The cosmic ray spectrum may be found at the Particle Data Group site (http://pdg.lbl.gov/) and is shown in Fig. 1 below.

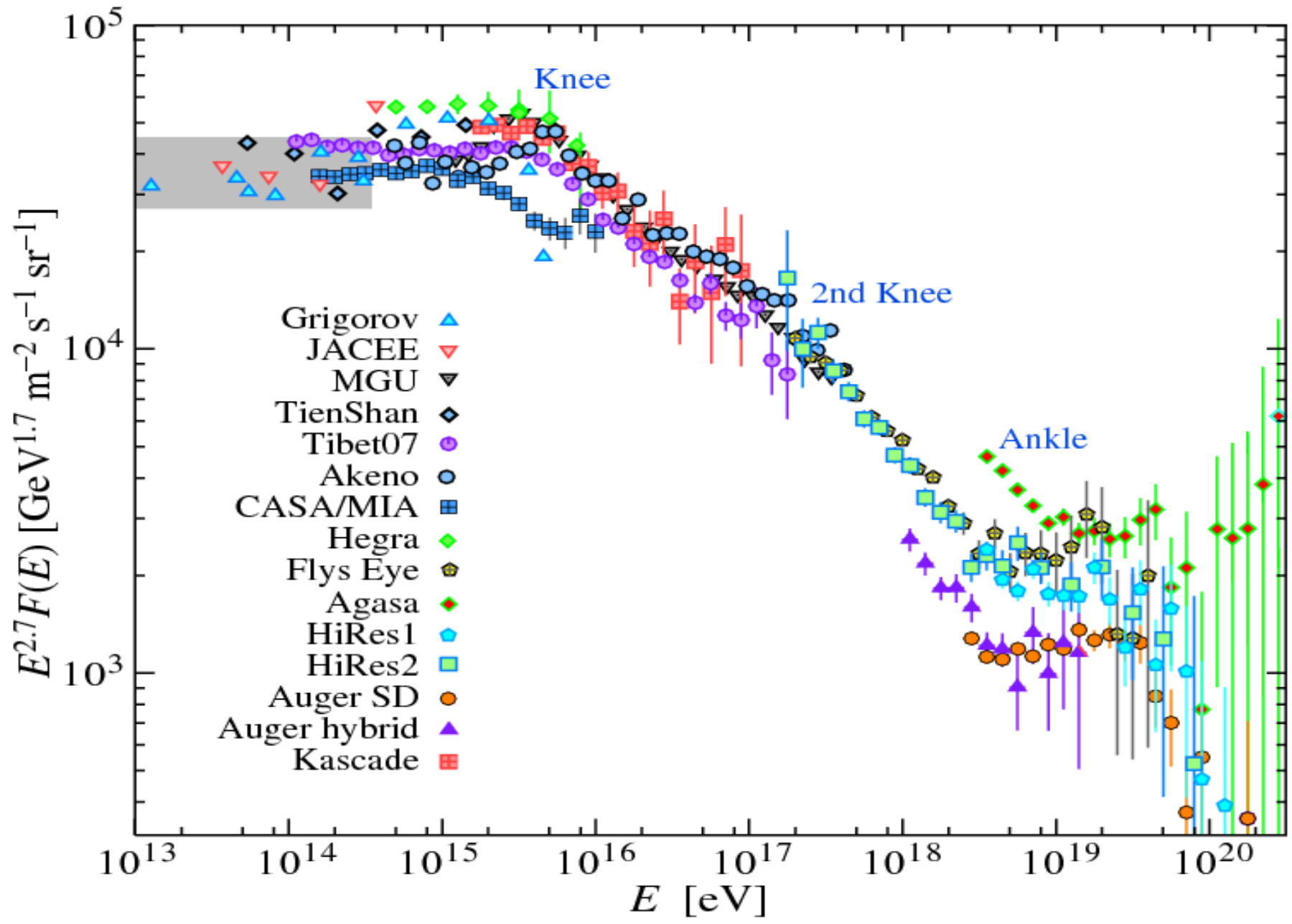

Figure 1: The cosmic ray flux distorted by a flattening factor of $E^{2.7}$ is indicated as a function of energy. It is assumed that most of the cosmic rays at high energies (> $10^{17}$ eV) are protons. The LHC-equivalent energy is $10^{17}$ eV, hence as the figure shows, near the second 'knee'.

## COSMIC RAY INTEGRATED FLUX FOR $E_C \geq 10^{17}$ eV

In order to estimate the integrated cosmic ray flux at and above the LHC-equivalent energy the data of Fig. 1 were converted to the functional form given in Eq. 3 below.

$$F(E) = \frac{df(E)}{dE}\left[\frac{1}{GeV\,m^2 sSr}\right]. \quad (3)$$

The integral of the function over energy therefore provides the high energy flux per [$m^2$ s Sr]. This is achieved by fitting and integrating a power law function F(E) = $2.6\times10^{7}\ E^{-3.1}$ determined from the data. Fig. 2 shows the power law fit superimposed on the data where it is found that:

$$\Phi = \int_{E_c \geq 10^{17} eV} \frac{df(E)}{dE}\, dE = 1.5\times10^{-10}\ m^{-2}\ s^{-1}\ Sr^{-1} \quad (4)$$

It is this flux that is integrated over the sun and earth during their lifetimes (~ 4 Gyr [1]) that will be compared to the LHC flux at design luminosity integrated over 10 years of operation.

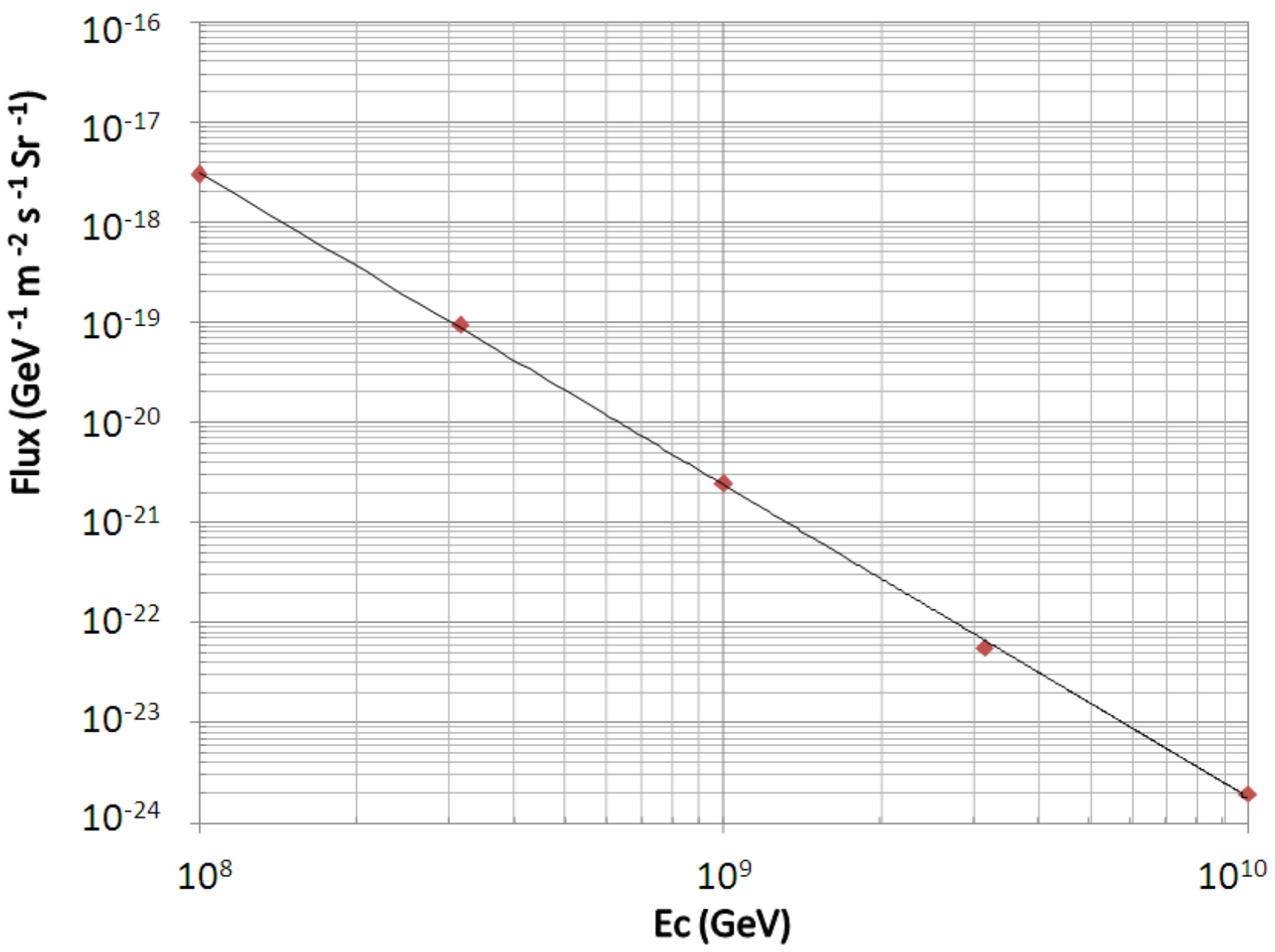

Figure 2: Five data points were used for an approximate integration of the cosmic ray flux from LHC equivalent energy to several orders of magnitude above. A power law fits the data quite well and is indicated by the line in the fig.

## COMPARISON OF COSMIC RAY VERSUS LHC PRODUCTION OF EXOTICS

Neither the cross section for the particle production of mini-black holes nor other exotics is known. For that matter, at this writing it is highly speculative whether or not these exotic objects even exist on the elementary particle scale. Nevertheless, it is posited that if such productions were possible, cosmic rays hitting the sun (earth) over its lifetime at the LHC equivalent energy and above can be compared with the production at the LHC. It is postulated that if a mini-black hole or other pernicious exotics were produced the sun (earth) would have imploded (or exploded), making in either case our earth's 4 Gyr existence unlikely.

---

[1] The age of the solar system is generally taken to be 4.55 Gyr based on radiometric dating. The integration time for cosmic rays in this analysis was taken to be 4 Gyr, about 14% shorter. It is assumed that the cosmic flux has been relatively constant over the age of the solar system even though the universe was more violent in earlier times.

Two calculations are made:

- 'Thick Target Assumption' – where the total number of cosmic rays impinging on the sun (earth) and interacting with nucleons within the radius of the sun's photosphere (earth) is compared to the total interactions at the LHC with an inelastic cross section $\sigma \sim$ 100 mb. The density of the photosphere is $\rho \sim 1.7\text{x}10^{-7}$ g/cm$^3$. Hence, one interaction length (51 g/cm$^2$) corresponds to $3\text{x}10^6$ m, or about 0.4% of the solar radius. The density profile of the sun increases with decreasing radius making the 'Thick Target' even a better assumption. The earth's atmosphere is about 1,000 g/cm$^2$ integrated down to sea level – hence it is a 'thick' target.

- 'Thin Target Assumption' – where it is assumed that a component of the cosmic ray flux is highly penetrating (neutrinos or exotics produced from the primary interaction) and thus the relevant measure is the nucleon column densities of the sun and of the earth. In these comparisons integrated luminosities are estimated - analogous to calculations performed for neutrino-nucleon interactions in a laboratory neutrino experiment.

The relevant data for these two comparisons are summarized in Table 1 below.

**TABLE 1**

| Quantity | Value | Units |
|---|---|---|
| CR luminosity $\sqrt{s} \geq 14$ TeV | $1.5\text{x}10^{-10}$ | m$^{-2}$ s$^{-1}$ Sr$^{-1}$ |
| Radius of sun | $6.9\text{x}10^{8}$ | m |
| Area of photosphere | $6.0\text{x}10^{18}$ | m$^2$ |
| Mass of sun | $2.0\text{x}10^{30}$ | kg |
| Nucleons in sun | $1.2\text{x}10^{57}$ | |
| Solar target density $\lambda_s$ | $1.3\text{x}10^{11}$ | g/cm$^2$ |
| Radius of earth | $6.4\text{x}10^{6}$ | m |
| Area of earth surface | $5.1\text{x}10^{14}$ | m$^2$ |
| Mass of earth | $6.0\text{x}10^{24}$ | kg |
| Nucleons in earth | $3.6\text{x}10^{51}$ | |
| Terrestrial target density $\lambda_e$ | $4.7\text{x}10^{9}$ | g/cm$^2$ |
| LHC luminosity | $10^{34}$ | cm$^{-2}$ s$^{-1}$ |
| LHC Operation time/year | $10^{7}$ | s |

Comparison of High Energy Interactions - Thick Target Assumption:

From the numbers in the table above the integrated number of cosmic ray interactions in the sun over 4 Gyr is estimated to be $7.1\text{x}10^{26}$ computed by assuming that every incident cosmic ray interacts with a hydrogen nucleus - in essence proton (cosmic ray) interacts with proton (sun). The number of interactions at the LHC (proton-on-proton) for 10 years of operation is $1.0x10^{17}$. Hence, there are ~ 10 orders-of-magnitude between the numbers of interactions of cosmic rays hitting the sun versus protons smashing protons at the LHC. Even if the LHC were operated at the proposed upgraded luminosity of $10^{35}$ $cm^{-2}$ $s^{-1}$ for 10 years, there are still 9 orders-of-magnitude difference.

Doing the same calculation for the earth the integrated number of cosmic rays for a 4 Gyr exposure is $6.0x10^{22}$. As in the case of the sun, since the earth target is 'thick' every one of these cosmic rays will interact. The safety factor in this case would be ~ $6x10^5$ or $6x10^4$ – depending on the LHC luminosity.

Thus, if 'pernicious' exotics can be produced at LHC energies, they would have already been made with high probability during the presently elapsed lifetime of the sun and earth by a very large factor. Even considering only the earth target, the 'safety factor' for experiencing a pernicious interaction is quite small ≤ $1.7x10^{-5}$.

Comparison of Luminosities - Thin Target Assumption:

The column density of the sun is estimated by taking the number of nucleons contained within the sun divided by the cross sectional area of a circle of radius of the photosphere. This calculation is relevant if there were a highly penetrating component, such as neutrinos, of the cosmic ray flux or if the initial cosmic ray interaction produced highly penetrating particles. Again, from Table 1, the column density of the sun is ~ $8x10^{34}$ nucleons/$cm^2$. The integrated cosmic ray flux over this circular area for 4 Gyr is ~ $3.6x10^{26}$. In the case of the earth, the integrated cosmic ray flux through its corresponding cross section for 4 Gyr is $3.0x10^{22}$. The column density of the earth is ~ $2.8x10^{33}$ nucleons/$cm^2$.

Hence the 'luminosity' of cosmic rays hitting the sun is $L_{sun}$ ~ $(8x10^{34}/cm^2)x(3.6x10^{26})$ ~ $2.8x10^{61}/cm^2$ and the corresponding number for the earth is $L_{earth}$ ~ $(2.8x10^{33}/cm^2)x(3.0x10^{22})$ ~ $8.4x10^{55}/cm^2$. These target densities are to be compared with the integrated luminosity (effective target density) of the LHC operating for 10 years which is computed to be $L_{LHC}$ ~ $10^{42}/cm^2$ (or $10^{43}/cm^2$ with the luminosity upgrade). Thus, there are 19 to 18 orders of magnitude 'safety' (for the earth 13 to 12 orders of magnitude) of the integral cosmic ray luminosity relative to the corresponding to the LHC design luminosity and the upgraded luminosity, respectively.

If production of pernicious exotics is possible at the LHC for 10 years of running the same process would have destroyed the sun and the earth in very brief time. Taking the earth as an example, cosmic ray production of pernicious exotics would have destroyed our planet in ~ 10 yr/$10^{12}$ ~ $10^{-4}$ s – in clear disagreement with known lifetime of the earth.

The results given above are summarized in Table 2. In the table cosmic rays (CR) are integrated over 4 Gyr and the LHC numbers are computed for the machine operating at the design luminosity (L ~ $10^{34}$ $cm^{-2}$ $s^{-1}$) for 10 years. Note that the ratio of cosmic rays to LHC operation for each measure is given in the right-most column of the table.

**TABLE 2**

| Thick Target | CR Total Interactions | LHC σ ~ 100 mb | Ratio CR/LHC |
|---|---|---|---|
| Sun | $7.1\text{x}10^{26}$ | $1.0\text{x}10^{17}$ | $7.1\text{x}10^{9}$ |
| Earth | $6.0\text{x}10^{22}$ | $1.0\text{x}10^{17}$ | $6.0\text{x}10^{5}$ |
| Thin Target | CR Luminosity | $L_{LHC}$ | |
| Sun | $2.8\text{x}10^{61}/\text{cm}^2$ | $10^{42}/\text{cm}^2$ | $2.8\text{x}10^{19}$ |
| Earth | $8.4\text{x}10^{55}/\text{cm}^2$ | $10^{42}/\text{cm}^2$ | $8.4\text{x}10^{13}$ |

**HAWKING RADIATION**

The famous Hawking mechanism [3] postulates that by quantum fluctuations of the vacuum at the event horizon, black holes will have a finite temperature – thereby radiating one half of the particles torn from the vacuum, while accreting the other (negative energy) half. Thus, in absence of external matter or radiation falling into it, black holes will have a finite lifetime – growing hotter with time to eventually evaporate in a burst of gamma rays. On an astronomical scale, a black hole of the mass of the earth will have a temperature of about 2.7 $^{o}$K and a lifetime of $5.7\text{x}10^{50}$ yr neglecting cosmic microwave background (CMB) and other external disturbances which would make the lifetime longer[2].

Using the Hawking formula for a black hole of mass 7 TeV/$c^2$ ($1.24\text{x}10^{-23}$ kg), the lifetime is ~ $1.6\text{x}10^{-85}$ s and is shorter than the 'classical' Planck time ($10^{-43}$ s) – hence unphysical. But with large extra dimensions, the 'Planck length scale' is theorized to be much larger making the lifetime of a black hole with the mass of a few TeV/$c^2$ of the order of the new (longer) Planck time. In models with large extra dimensions, the lifetime scales as:

$$t \sim \frac{1}{M*} \left(\frac{M_{bh}}{M*}\right)^{(n+3)/(n+1)} , \qquad (5)$$

where M* is the mass scale of the large extra dimension, $M_{bh}$ is the mass of the mini-black hole and n is the number of extra dimensions.

We consider a black hole of $M_{bh} \sim \sqrt{s}/2 = 7$ TeV/$c^2$. Taking a theory of worse case with only one extra dimension, n = 1, and the new Planck scale M* = 1 TeV, we find the lifetime of the mini-black hole to be t ~ $3.3\text{x}10^{-26}$ s. (Note 1 $\text{TeV}^{-1}$ ~ $6.7\text{x}10^{-28}$ s.) Hence, the min-black hole will not live beyond the microscopic scale at which it is produced. To give an order of magnitude comparison, the transit time of a photon at c across 1% of a proton diameter is of the same order. In the case of cosmic rays with a Lorentz boost of approximately ~ $10^4$ from the COM frame to the lab frame, the mini-black hole will not live very long or travel very far. For completeness, a black hole of 14 TeV/$c^2$, the mass at the kinematic limit of the LHC, would have a lifetime 4 times longer than the 7 TeV/$c^2$ one.

---

[2] A black hole at 2.7 $^{o}$K is in thermal equilibrium with the present microwave background radiation and thus would absorb photonic radiation as much as it would radiate. The universe would have to become cooler before the evaporation process starts to be effective. The Schwarzschild radius of an earth-size black hole is ~ 8.9 mm.

## BETHE-BLOCK ENERGY LOSS

It is interesting to consider the case when the exotic does not evaporate quickly by Hawking radiation but instead continues to travel through matter, 'eating' mass as it goes. Such a situation would exist for negatively charged strange matter theorized by Farhi and Jaffe [5].

Assume that an exotic particle has a mass $M_s c^2 \sim \sqrt{s} \sim 14$ TeV and is produced in a proton-proton collision at threshold. If this collision occurred at the LHC, the exotic would be at rest in the lab frame, whereas if it were produced by an energetic cosmic ray hitting a stationary proton it would be Lorentz boosted in the lab frame with $\gamma \sim p^*/M_p \sim 7.5 \times 10^3$ (γ-factor of the COM frame) and have a lab momentum approximately equal to that of the incident cosmic ray ~ 1 x $10^5$ TeV = $10^{17}$ eV.

One might conclude that the exotic would completely penetrate the sun and thus the cosmic ray calculations discussed above would not be a relevant comparison to the hypothesized process at the LHC where the exotic particle may be produced at rest. It is argued that the cosmic ray comparison is still cogent because the exotic will probably have an electric charge, such as in the case of a strangelet. When produced by cosmic rays striking the sun, the charged exotic will lose energy by interacting with the electrons of the solar medium.

In order to estimate the stopping power of the sun the Bethe and Block[3] expression is used [6]. Assuming that the exotic particle has a charge of z = 1 (strangelets will probably have a larger charge) and dE/dx ~ 4 MeV/(g/cm$^2$) (minimum ionizing in hydrogen) we compute the average energy loss of a particle traversing the sun to be

$$\Delta E \sim \int \frac{dE}{dx}\, dx \sim 1.3 \times 10^{11}\ \mathrm{g/cm^2} \times 4.1\,\mathrm{MeV/(g/cm^2)} \sim 5.2 \times 10^{17}\ \mathrm{eV}. \qquad (6)$$

Hence, a $1 \times 10^{17}$eV strangelet would range-out (be stopped) in the sun. If it has pernicious business to do, such as eating matter, it would have had the opportunity to do so.

## CONCLUSIONS

Comparing the cosmic ray interactions at and above the equivalent LHC energy indicates that pernicious exotics (mini-black holes, etc.) would have occurred with much higher probability over astronomical time than by the operation of the LHC by many orders of magnitude. In addition there are two other factors against unique production of pernicious exotics happening at the LHC. If mini-black holes were to be produced they would decay in microscopic distances from the beam-beam interaction point. Strangelets or other charged exotics would have ranged out in the sun and if produced would have had ample opportunity to do mischief over 4 Gyr [7]. The lifetimes of the sun and earth preclude this as a possibility. The highly imaginative may continue to speculate – such as postulating that the LHC will produce *LHConium,* a pernicious exotic, under conditions not probed by cosmic rays, but the probability of such a thing is extremely small[4].

[3] The argument presented here is a simplification – but its essence is that the exotic produced by a cosmic ray – proton collision will lose energy by interacting with the matter of the sun. There is nothing unique about the proton-proton collisions at the LHC. The physics of energy loss of charged particles moving through plasma is of interest in nuclear fusion reactors, plasma accelerators as well as astrophysics.

## APPENDIX

For completeness the data of the very upper range of the measured cosmic ray spectrum used the PDG compilation is shown in Fig. A1.

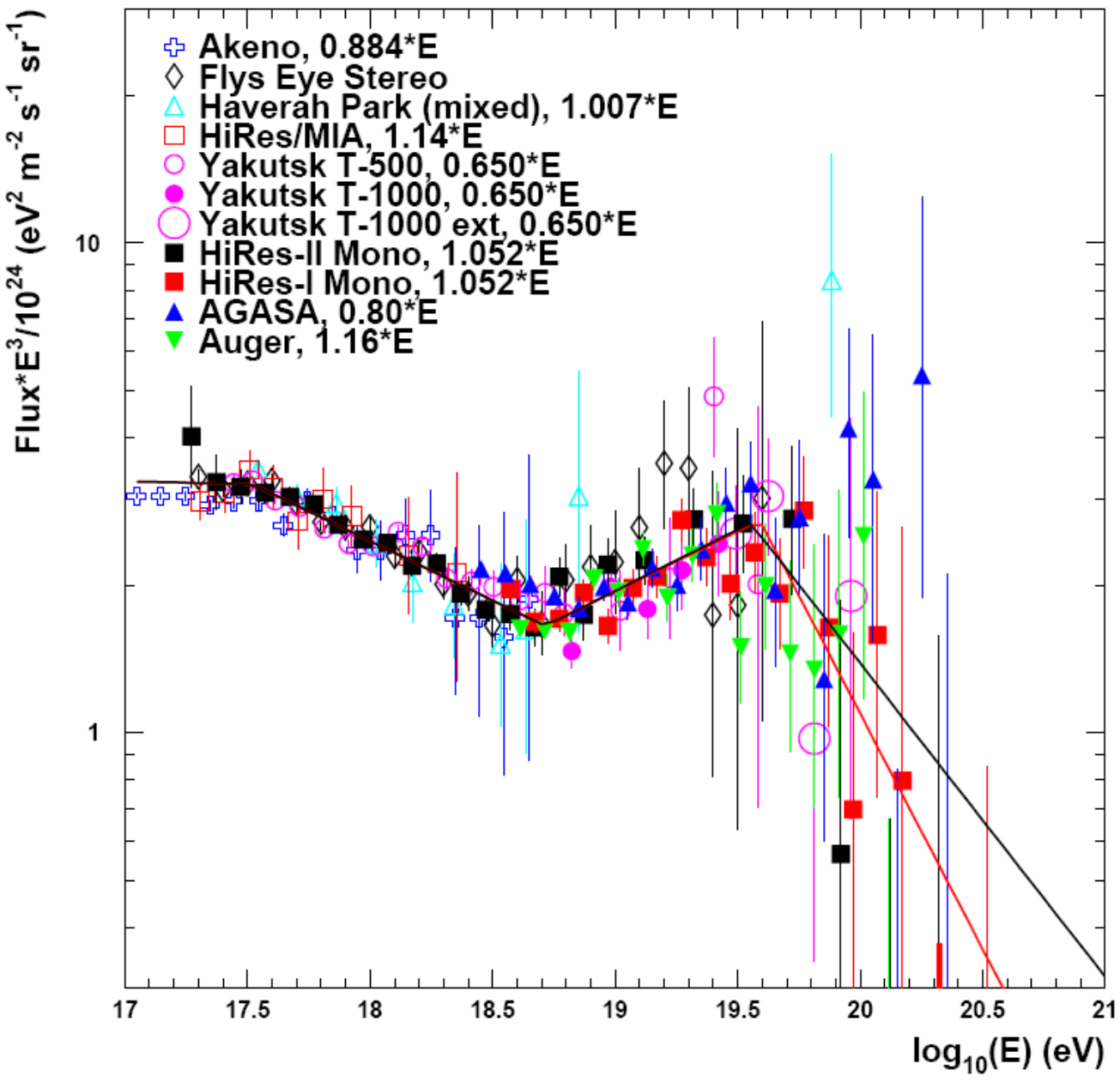

Figure A1: The very high end of the measured cosmic ray spectrum is shown (D. R. Bergman and John W. Belz (arXiv:0704.3721v1 [astro-ph] 27-Apr-2007). The black line is a fit with the AGASA data and red without.

---

[4] The author has sufficient confidence in the non-threat of the LHC operation that he offers to buy all of his friends bottles of *Mouton Rothschild 1945* if the conclusion of this note is incorrect – of course assuming adequate supply of this rare wine exists after the celebrations of the LHC startup.